\documentclass{elsart}

\usepackage{natbib, graphicx}

\usepackage{amssymb}
\journal{Physics Letters A}
\begin{document}

\begin{frontmatter}



\title{\bf From particle segregation to the granular clock}

\author[brenig]{R. Lambiotte},
\ead{rlambiot@ulb.ac.be}
\author[marcos]{J.M. Salazar}
\ead{jmarcos@u-bourgogne.fr}
\author[brenig]{L. Brenig}
\ead{lbrenig@ulb.ac.be}

\address[brenig]{Physique Statistique, Plasmas et Optique Non-lin\'eaire, Universit\'e Libre de Bruxelles, Campus Plaine, Boulevard du Triomphe, Code Postal 231, 1050 Bruxelles, Belgium}
\address[marcos]{LRRS UMR-5613 CNRS,  Facult\'e des Sciences Mirande, 9 Av. Alain Savary 21078 Dijon Cedex , France. }

\begin{abstract}
Recently several authors studied the segregation of particles for a system composed of mono-dispersed inelastic spheres contained in a box divided by a wall in the middle. The system exhibited a symmetry breaking leading to an overpopulation of particles in one side of the box. Here we study the segregation of a mixture of particles composed of inelastic hard spheres and fluidized by a vibrating wall. Our numerical simulations show a rich phenomenology: horizontal segregation and periodic behavior. We also propose an empirical system of ODE's representing the proportion of each type of particles and the segregation flux of particles. These equations reproduce the major features observed by the simulations.  

\end{abstract}

\begin{keyword}
Granular segregation \sep Granular flow \sep Molecular Dynamics simulations 
\PACS 45.70.Vn\sep 45.70.Mg

\end{keyword}
\end{frontmatter}

\section{Introduction}

\begin{figure}
\includegraphics[width=1.35in]{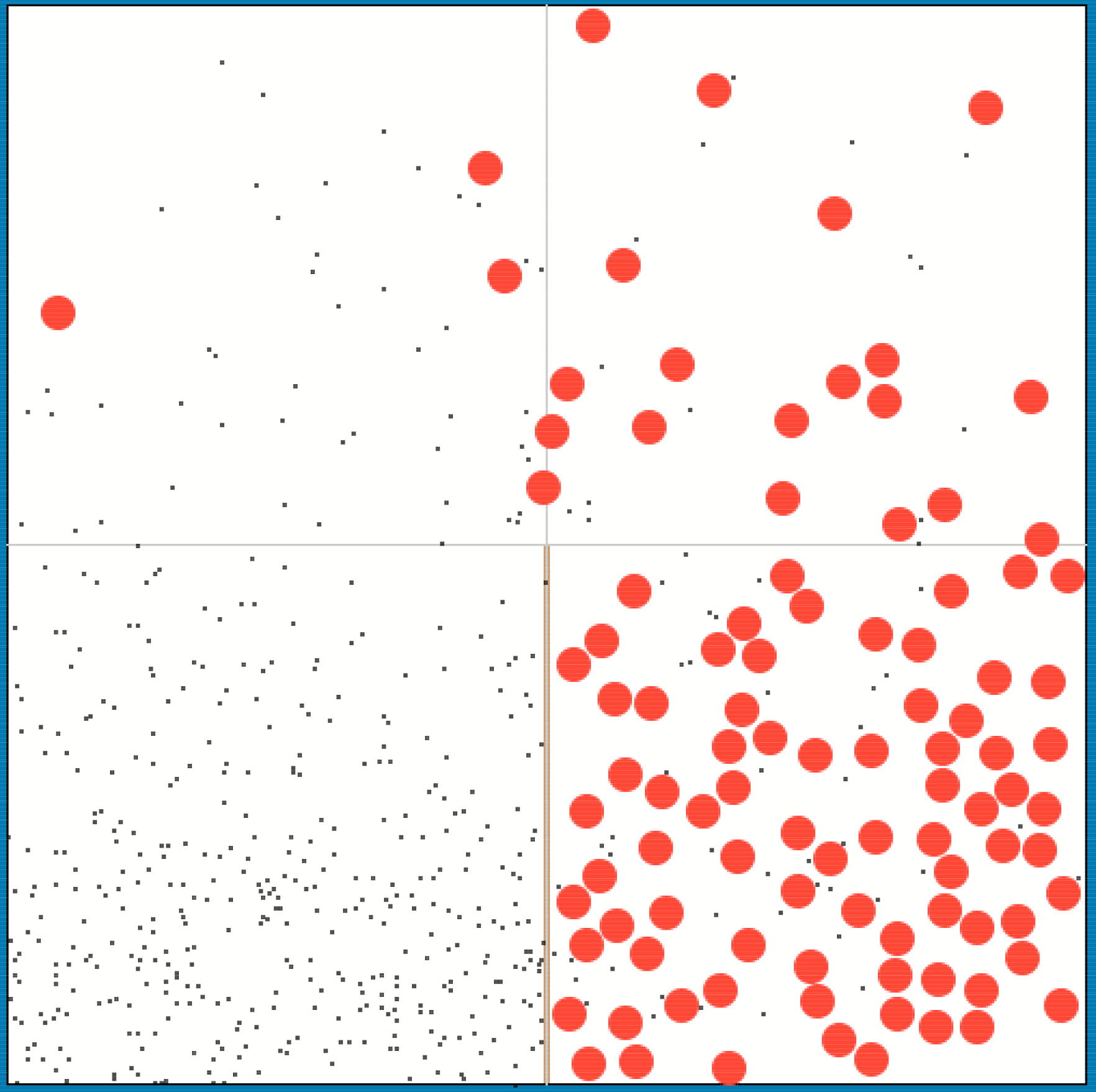}
\includegraphics[width=1.35in]{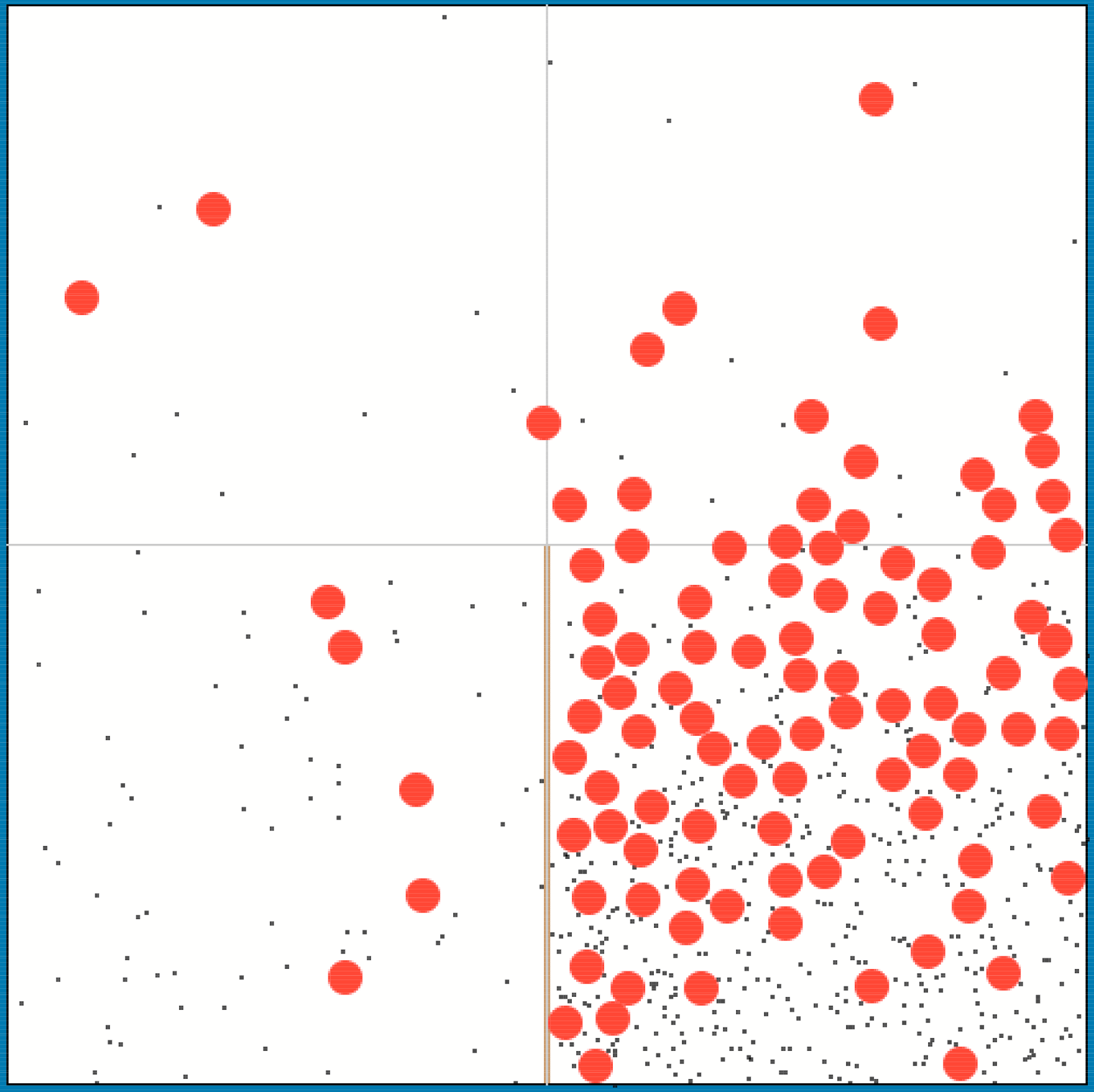}
\includegraphics[width=1.35in]{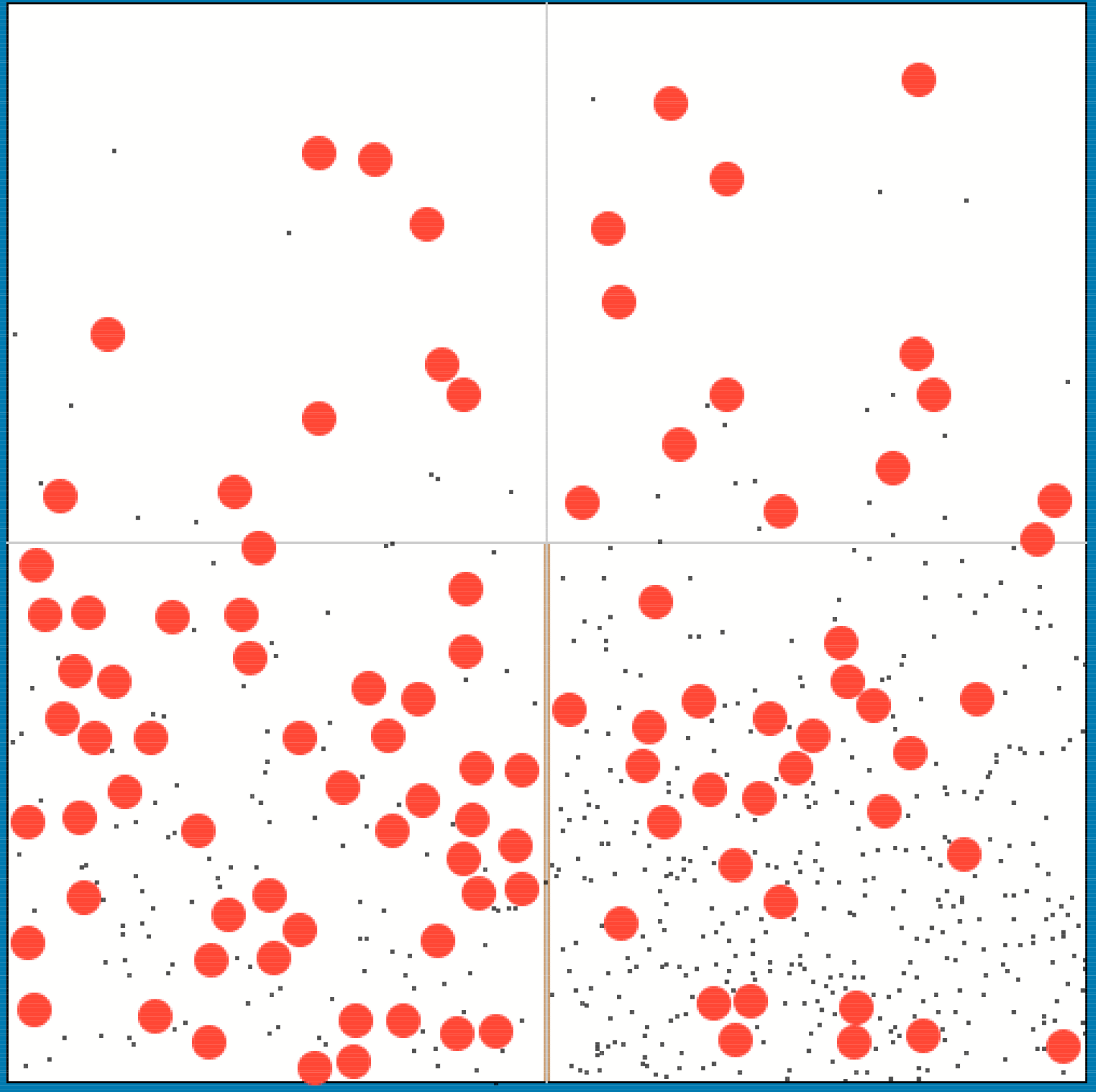}
\includegraphics[width=1.35in]{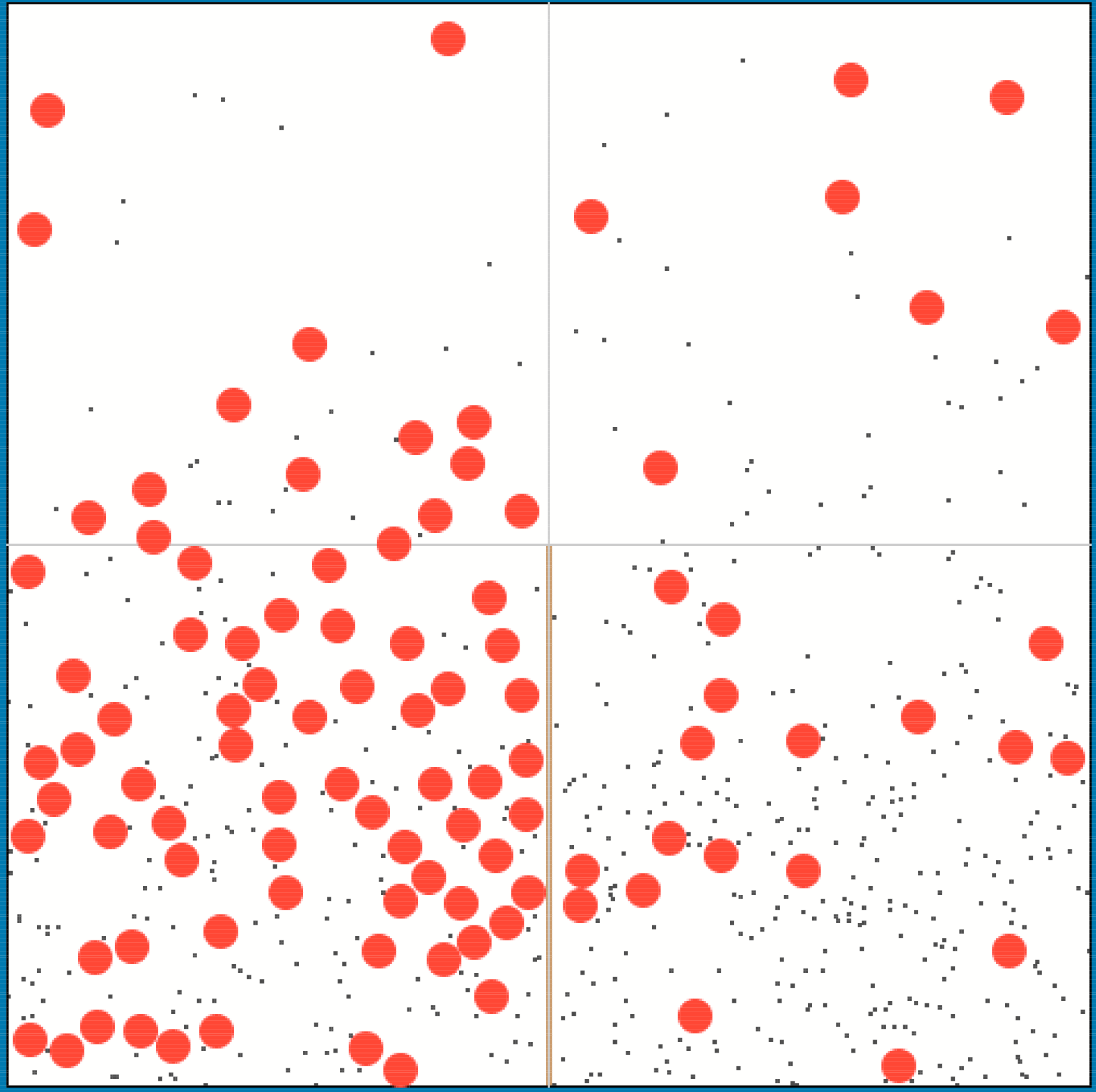}
\caption{\label{figure1} (a),(b),(c),(d) Snapshots of the oscillation at different times. In (a), one plots the initial condition, with the small discs in box L and the large discs in box R. In (b), one may observe the transfer of small discs into box 2 and the vertical segregation. In (c), the transfer of large discs into box L has started. In (d), most of large discs are in box L.  }
\end{figure}

Granular fluids are assemblies of moving macroscopic solid
grains,  including common materials such as; sand, powders, cereals, etc. Their ubiquity in various industrial processes (e.g., pharmaceutical, food) as well as in geophysics  and astrophysics motivates a better understanding of their properties \cite{intro}. Due to  the inelastic interaction between the grains, these systems have a natural tendency to cool down  by successive dissipative collisions leading to a macroscopic state of rest. Nevertheless, when these systems are supplied with an external energy source (e.g., a vibrating wall), they reach a dynamic stationary state  whose macroscopic behavior  presents some similarities with a classical liquid or a gas. Moreover, granular fluids exhibit a rich and peculiar phenomenology for instance; non-gaussian velocity distributions, non-equipartition of energy, anomalous transport coefficients, a natural tendency to  form  dense and dilute regions \cite{gold}, shearing instability, etc. Recently, a significant progress has been achieved in the understanding of these phenomena by applying the kinetic theory and the resulting hydrodynamical equations.  Some years ago a generalized Maxwell demon experiment \cite{daemon} has been devised that shed light on one of the most amazing properties of granular media. This experience  consists of a box divided into two equal compartments by a vertical wall at the bottom and going up to the middle of the box, leaving the upper half of the box undivided.  The box is filled with identical inelastic particles submitted to gravitation and the energy is supplied  by a vibrating wall (Fig.~\ref{figure1}). For a high energy input, the system presents a  homogeneous steady state, while when the energy input is decreased, a phase transition occurs and an asymmetric steady state prevails. 
This asymmetric steady state has been first studied theoretically by Eggers \cite{eggers} in a slightly modified configuration with a separation wall extending throughout the entire height of the box. This wall contains a small hole at a given height through which particles can be exchanged. In this formulation was assumed that the exchange of particles between the compartments corresponds to an effusive process and it was considered that the mean energy of the granular system decreases when the density increases. This last feature of granular systems is a consequence of the inelasticity of the collisions between particles and is responsible for the symmetry breaking in the Maxwell demon experiment. Indeed, an asymmetric fluctuation in the system induces a decrease of the mean energy in the denser compartment. Consequently, the escaping rate of grains from this compartment is reduced and the grains have higher  tendency to flow from the dilute compartment towards the dense compartment. This argument  has been successfully verified by both molecular dynamics  simulations \cite{deele} and hydrodynamic approaches \cite{brey}.  

In this work, we  focus on a Maxwell demon simulation with the same experimental setting as described in the first paragraph (Fig.~\ref{figure1}) and characterized by two equal components. The binary system is composed by two kinds of smooth hard discs, whose dynamics depends on a large number of parameters. Here we have assumed that the particles are characterized by the diameters $\sigma_{1}$, $\sigma_{2}$ and by the same mass ($m=1$). Concerning the  restitution parameter $\alpha$ it was set equal to $0.95$. The particles evolve under gravity, $\bf{g}$ (which acts along the y-direction), in a  rectangular two-dimensional box (Fig.~\ref{figure1}) of height H and width W, and moving ballistically between successive collisions. The collision rule  between grains i and j reads:  
\begin{eqnarray}
{\bf v}_{i}^{'} = {\bf v}_{i} -  \frac{1}{2}  (1+ \alpha ) {\mbox{\boldmath$\epsilon$}}
({\mbox{\boldmath$\epsilon$}}.{\bf v}_{ij}) ~~~~
{\bf v}_{j}^{'} = {\bf v}_{j} + \frac{1}{2}  (1+ \alpha ) {\mbox{\boldmath$\epsilon$}}
({\mbox{\boldmath$\epsilon$}}.{\bf v}_{ij}) 
\end{eqnarray} 
with ${\bf v}_{ij} \equiv {\bf v}_{i} - {\bf v}_{j}$ and ${\bf r}_{ij} \equiv {\bf r}_{i} - {\bf r}_{j}$.
The primed velocities are the post-collisional velocities and the unprimed ones  the velocities before the collision. ${\mbox{\boldmath$\epsilon$}} \equiv \frac{{\bf r}_{ij}}{|{\bf r}_{ij}|}$
is the unitary vector along the axis joining the centers of the two colliding
spheres.  These rules have been often used to study fluidized granular materials and  have been proven to reproduce the specific phenomenology of granular fluids. The equations of motion of the N particles are solved numerically using the event-driven algorithm \cite{Luding}. In our simulations, the box is divided in two equal compartments, box L (left) and box R (right) of width W/2, by a wall parallel to the y-direction of height H/2. The walls are elastic except the ground which is a thermalizing wall at a given temperature $T_G$ injecting the energy to the system.    For this we used the dimensionless parameter $\gamma \equiv (\frac{T_{G}}{m g H})$  relating the gravity and the energy input. The system studied is composed by 500 {\em small} particles and 100 {\em big} particles. The diameter ratios analyzed were 8.0 and 4.0 and in both cases we obtained the same dynamical scenario as we will show below. It is important to point out that this way of injecting the energy into the system has no  intrinsic frequency and does not introduce any essential differences when compared to an injection of energy by a vibrating wall.

\section{Dynamical regimes}
\begin{figure}
\includegraphics[angle=-90,width=2.5in]{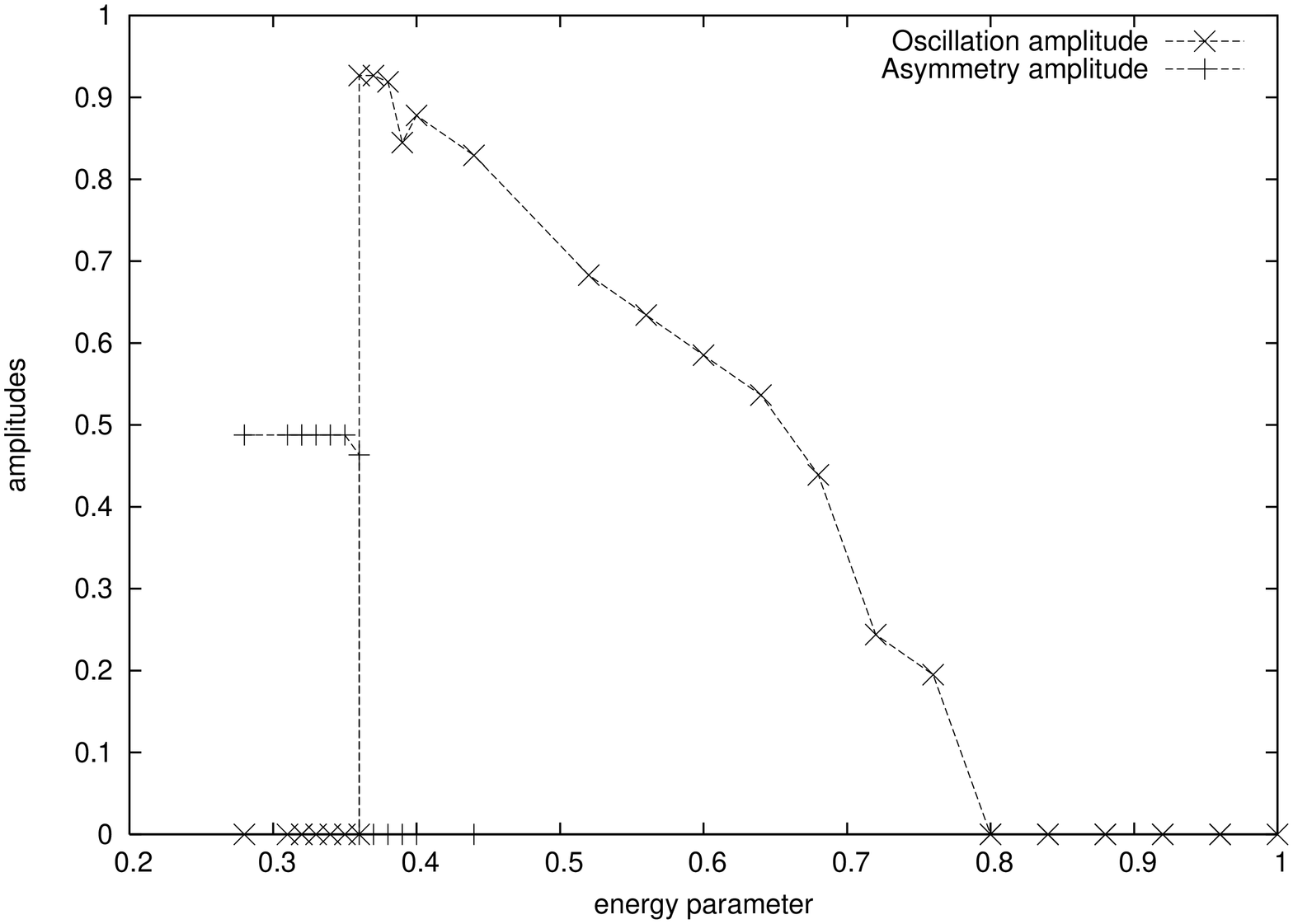}
\includegraphics[angle=-90,width=2.5in]{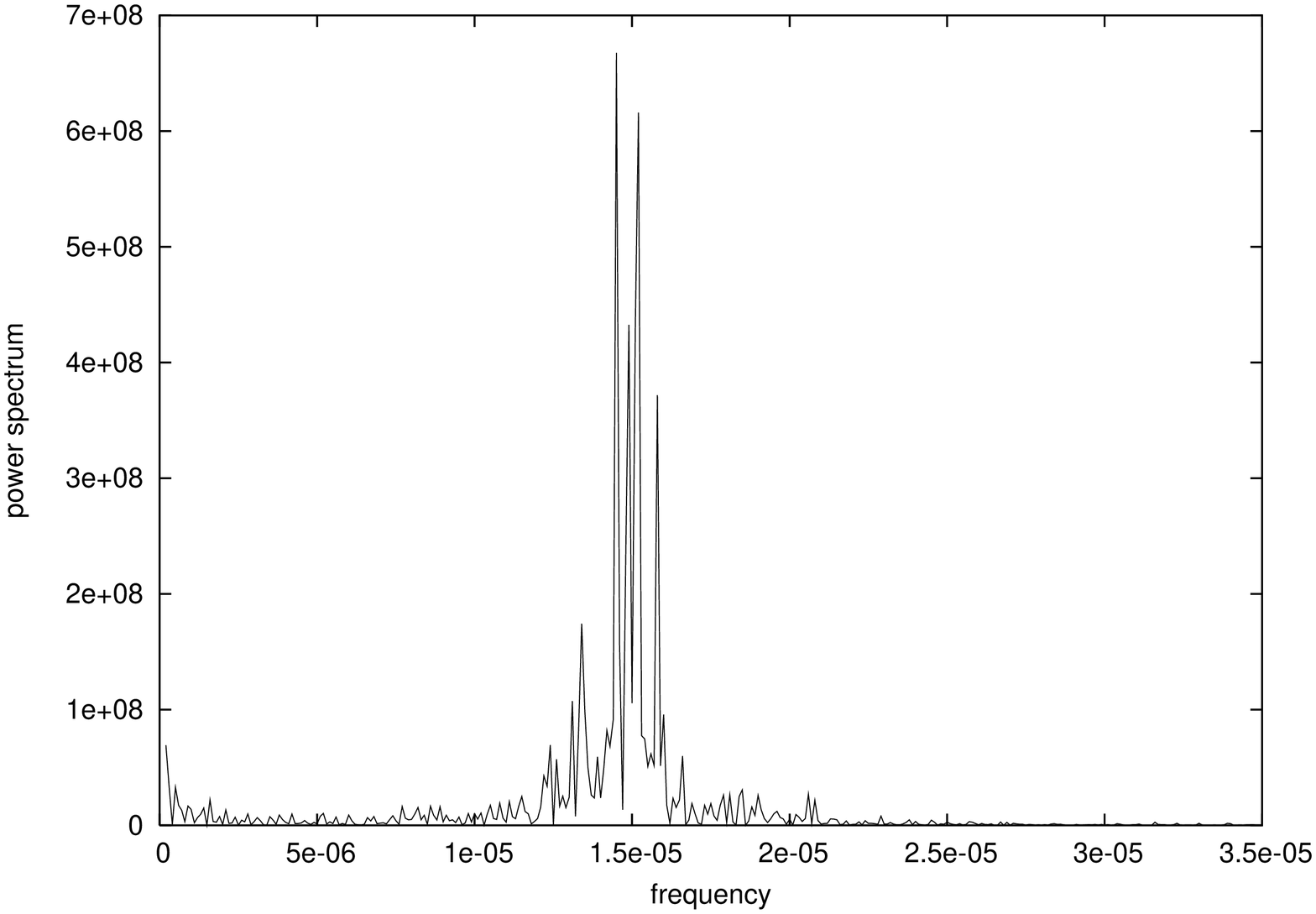}

\caption{\label{figure2a} In the left figure, bifurcation diagram obtained from event-driven simulations, as a function of the energy parameter $\gamma$. The amplitudes are the oscillation amplitude, characterizing the oscillation regime and an asymmetry amplitude, defined by $A_{Asym} \equiv <| n(t) - \frac{1}{2}|>$, which characterizes the asymmetric regime. In the right figure, discrete Fourier Transform of the autocorrelation of the signal in the periodic regime ($\gamma=0.48$) calculated for long runs ($5 \times ~10^8$ collisions). }
\end{figure}

 In the one component case the dynamical system is described by the proportion of particles on the left $n^L(t) \equiv \frac{N^{LEFT}(t)}{N}$. It is known that when  $\gamma$ is decreased  a transition occurs \cite{deele}. That is,  for $\gamma > \gamma^{*}$ on  average the particles distribute themselves uniformly over the two compartments: $n^L(t)  = 1/2$ and for $\gamma< \gamma^{*}$, one of the compartments gets denser, leading to an asymmetric stationary solution $n^L(t)  \neq 1/2$. This symmetry breaking is entirely due to the inelastic coupling between density and temperature explained in the introduction. For a binary granular fluid other key features come into play. Indeed, when granular media are shaken, particles of different sizes tend to separate vertically \cite{brazil},\cite{hong}. Moreover, the system exhibits non equipartition of energy (i.e., the mean energy of the different species is different \cite{wild}) and as we will see below this is reflected by a difference in the granular temperature for each kind of particle. 
The coupling of these phenomena with the horizontal spatial instability described above leads to a surprising new phenomenon,  namely the emergence of an oscillatory regime.
In the following, the state of the system is characterized by two dynamical variables: the proportions of small and big particles on the left $n_{1}^{L}(t)=\frac{N_{1}^{LEFT}(t)}{N_{1}}$ and $n_{2}^{L}(t)=\frac{N_{2}^{LEFT}(t)}{N_{2}}$ respectively.   In our simulations we identified three well-defined regimes, separated by the critical values $\gamma^{*}$ and  $\gamma^{**}$, with $\gamma^{*} > \gamma^{**}$ (Fig.~\ref{figure2a}):

-{\em Stationary homogeneous regime ($\gamma > \gamma^{*}$):}  If the shaking of particles is vigorous enough, the particles are uniformly distributed over the two compartments. In this regime, which has been also observed for a case of a monodispersed distribution of particles,  the system   homogenizes like a classical elastic fluid:  $n_{1}^{L}(t) Å1/2$ and $n_{2}^{L}(t) Å1/2$.  The stability of this solution has been corroborated by simulations with different initial conditions.

\begin{figure}

\includegraphics[angle=-90,width=2.5in]{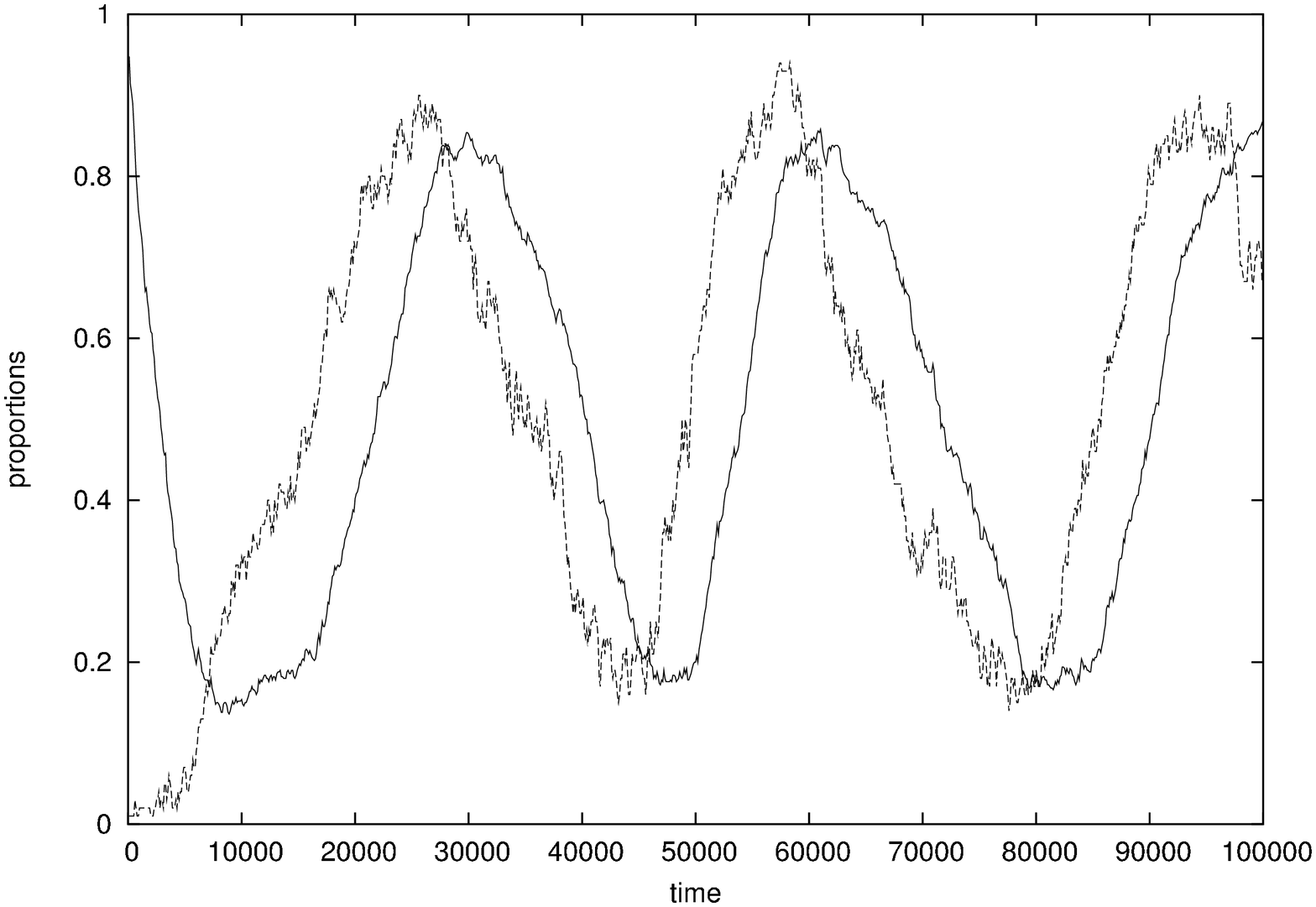}
\includegraphics[angle=-90,width=2.5in]{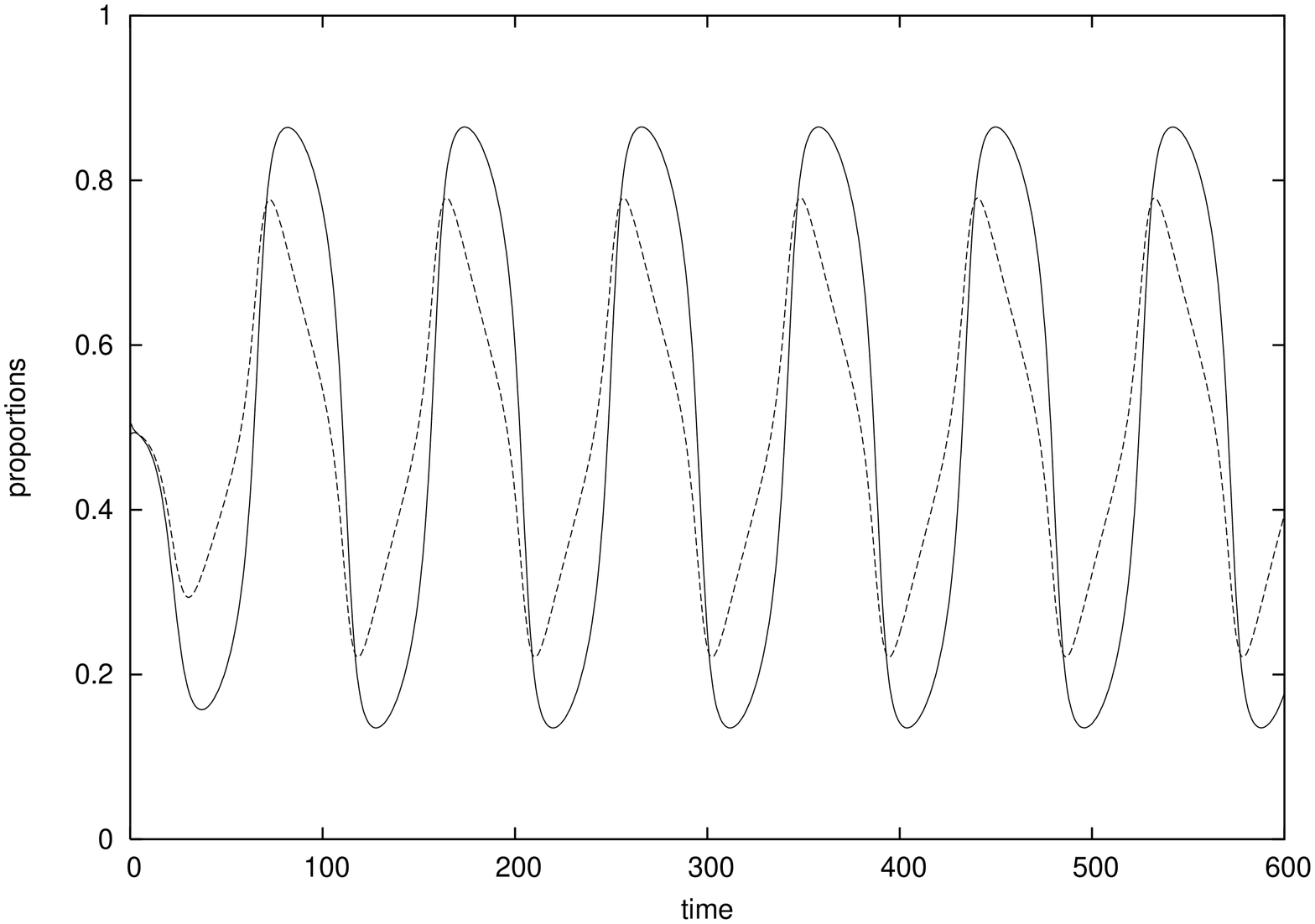}

\caption{\label{figure2b} Typical time evolution of the densities in the periodic regime respectively  for the simulations ($\gamma=0.58$) and for the proposed model ($\lambda$=0.1, R=3.0, A=1.67). }
\end{figure}

-{\em Periodic regime  ($\gamma \leq  \gamma^{*}$)}: When $\gamma$ is decreased below the critical value $\gamma^{*}$, i.e. when the energy input is lowered, a new behavior appears: the two densities of particles evolve periodically in time with the same frequency and separated by a phase delay (Fig.~\ref{figure2b}). This is what we call a granular clock. The mechanism behind these oscillations is the following. Suppose one starts the experiment with the small discs in box L and all the large discs in box R (Fig.~\ref{figure1}.a). Initially, the net flux of the small discs is directed towards the compartment of the large discs (Fig.~\ref{figure1}.b) since the latter suffer more collisions (the collision frequency  is proportional to the diameter) and dissipate more energy, favouring the clustering instability. Progressively, the small discs accumulate in the box R and consequently a vertical segregation takes place. Our simulations show that  the large discs are pushed upwards in the box (Fig.~\ref{figure1}.c). This vertical asymmetry, in the box R, increases the probability of escaping of the large discs  and diminishes the probability of escaping of the small discs. This will induce a flow of large discs towards the box L, while small discs continuing to flow from  L towards  R. Hence, most of the large discs finish in box L and  box R becomes rich on small discs (Fig.~\ref{figure1}.d). This situation is the opposite of the initial condition. Then, the whole process reverses and the initial state is recovered by the same mechanism. Hence, one oscillation has been taken place. Obviously, the motion of large discs drives the oscillation and this may  explain the observed phase delay  between the two signals. The power spectrum for long time runs of this regime consists of  a well-defined peak (Fig.~\ref{figure2a}) surrounded by smaller peaks. When $\gamma$ is decreased below the threshold $\gamma^{*}$,  the size of the small peaks increases with respect to the main peak. The fact that this effect increases with the distance to the  bifurcation point seems to indicate that  some non-resonant intrinsic nonlinearities start to play a relevant role. This property has been also observed in the solutions of the model proposed in the next section. Finally, it is important to point out that in our simulations the mean energy of the large discs is lower than the small ones, as reported by Wildman and  Parker for a binary system of grains \cite{wild}.

-{\em Stationary asymmetric segregated regime ($\gamma <  \gamma^{**}$)}: Finally, when the input energy  is lowered further a stationary asymmetric state appears characterized by a horizontal symmetry breaking. Initially, the system behaves  as in the periodic regime but in this case the input energy is reduced  and  the number of large discs which escape from the box R is not large enough to launch the oscillation. As long as the vertical segregation takes place, small discs get trapped below the large discs in box R, in such a way that only large discs can flow from box R to box L. Hence, an asymptotic steady state is reached in wich box L is mainly composed of large discs. It is interesting to mention that Barrat and Trizac \cite{barrat} have already observed such stationary asymmetric state for a binary Maxwell demon simulation without gravity.

\noindent 

The succession of the above regimes is summarized in the bifurcation diagram of Fig. \ref{figure2a}. In addition to the regimes described above we also observed a transition regime characterized by intermittent oscillations between the two opposite steady asymmetric steady states. This regime, which appears near the transition between the periodic and the asymmetric states, may be associated to the action of the internal noise on the dynamics of the macroscopic variables near the phase transition. One should note that the intermittent regime complicates the study of the second transition and does not allow a clear determination of the system frequency at that point. Nonetheless, results obtained from the theoretical model proposed in the next section seem to indicate that the frequency at the transition point goes abruptly  to zero.

\section{Model}

In this section we introduce a simple empirical system of dynamical equation that reproduces qualitatively well the  regimes observed in the above described simulations. It formulation has been inspired by a model proposed by J.Eggers \cite{eggers} for the Maxwell Demon experiment with only one of disc diameter composition.  By using a hydrodynamic approach Eggers was able to find, under certain approximations, an expression for the escaping rate from one compartment  in terms of the proportion of particles in the considered compartment only. If $n$ denotes the proportion of particles, in say the left compartment, then its time evolution can be written as:
 $ 
\frac{\partial n}{\partial t} =  - nF(n)  + (1-n)F(1-n)
 $
 , where F(n), the escape rate per particle, is a decreasing function for large values of n. This feature represents the typical granular effect that in high density regions, the average kinetic energy is reduced by the high number of inelastic collisions, leading to a reduction of the escaping rate.
 In the right hand side appears a loss term,  the escaping flux, and a gain term which has
 the same functional structure as the gain term but in this case it depends on (1-n), the proportion of particles in the right compartment. This last point guarantees 
 the existence of a steady homogeneous solution: $n=1/2$. 
 In fact, the steady state condition leads to three stationary solutions: the homogeneous one and two asymmetric configurations that are in good quantitative agreement with numerical
 simulations and these results are robust for a large class of functions $F(n)$. 

Concerning the two-component problem described in the last section, we are looking for a model, as simple as possible, that can be able to reproduce the sequence of regimes obtained when decreasing the energy input, namely
the homogeneous steady state, the oscillating regime and the asymmetric steady state. 
Given the inherent complexity of the system, we will only require in this work a qualitative agreement with the simulations. For doing this we  define  two  variables in the left  compartment; the proportions of small, $n$, and large, $m$ particles. The corresponding proportion in the right half-box are $(1-n)$ and $(1-m)$ respectively. Here we assumed that the escape rate of the grains  from the left half-box is of the form, $ F(n,m) = e^{-\frac{ (n^{2} + R m^{2})}{A}}$. Where A is a positive parameter representing the balance between energy input and  inelasticity of collisions. This parameter decreases with the inelasticity  ($A=\infty$, for elastic collisions).  The positive parameter R describes the fact that the cross section of the big particles  is larger than for the small ones ( the cooling effect of the big particles is larger).

 The vertical segregation phenomena is taken into account, qualitatively, by introducing two more dependent variables  $s_L$ and $s_R$. They describe the influence of segregation on the escaping rate in the left and right compartments respectively. We assume that these variables change the escape rate per particle say in box L, by reducing this rate for the small particles, and by increasing it for the large particles, in agreement with the discussion of the previous section.
 By construction, the {\em segregation variables} should vanish when there is no vertical segregation. The next step consists in assuming that these variables reach respectively the values  $n m$ and $(1-n)(1-m)$ asymptotically. This expression qualitatively reproduces  the fact that there should be no segregation if a box is composed by only one component ($n=0$ or $n=0$), while segregation effects should be maximum if there are many small discs  and large discs in that box, for instance ($n=1$ and $m=1$). Before going further, one should insist on the fact that  the two supplementary variables $s_L$ and $s_R$ have no direct physical interpretation, they are formal variables used in order to mimic the effects of segregation on the system dynamics. However, these variables are necessary in order to reproduce the observed phenomenology. Indeed, a model where $s_L$ and $s_R$ are not included does not lead to an  oscillating regime. 
 By considering all the points described in this paragraph the modeling equations can be written as:

 \begin{eqnarray} 
 \frac{\partial n}{\partial t} &=&  - nF(n,m) (1-s_{L}) + (1-n)F(1-n,1-m) (1-s_{R}) \cr
 \frac{\partial m}{\partial t} &=&  - m F(n,m) (1+s_{L}) + (1-m)F(1-n,1-m) (1+s_{R}) \cr
 \frac{\partial s_{L}}{\partial t} &=&  - \lambda   (n m - s_{L}) ~~~~
 \frac{\partial s_{R}}{\partial t} =  - \lambda   ((1-n)   (1-m) -  s_{R})
\label{model}
  \end{eqnarray} 

\begin{figure}
\includegraphics[angle=-90,width=2.6in]{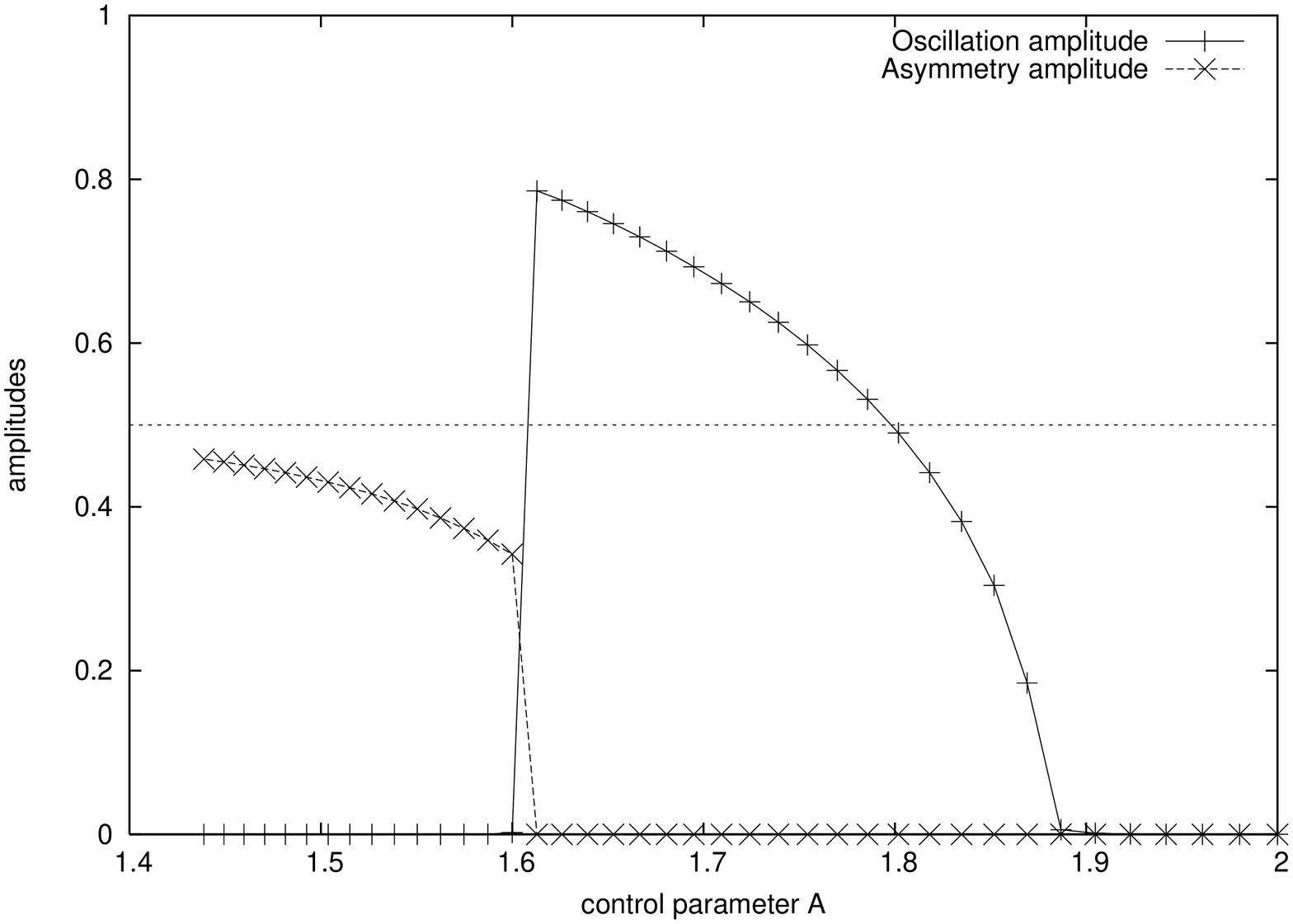}
\includegraphics[angle=-90,width=2.6in]{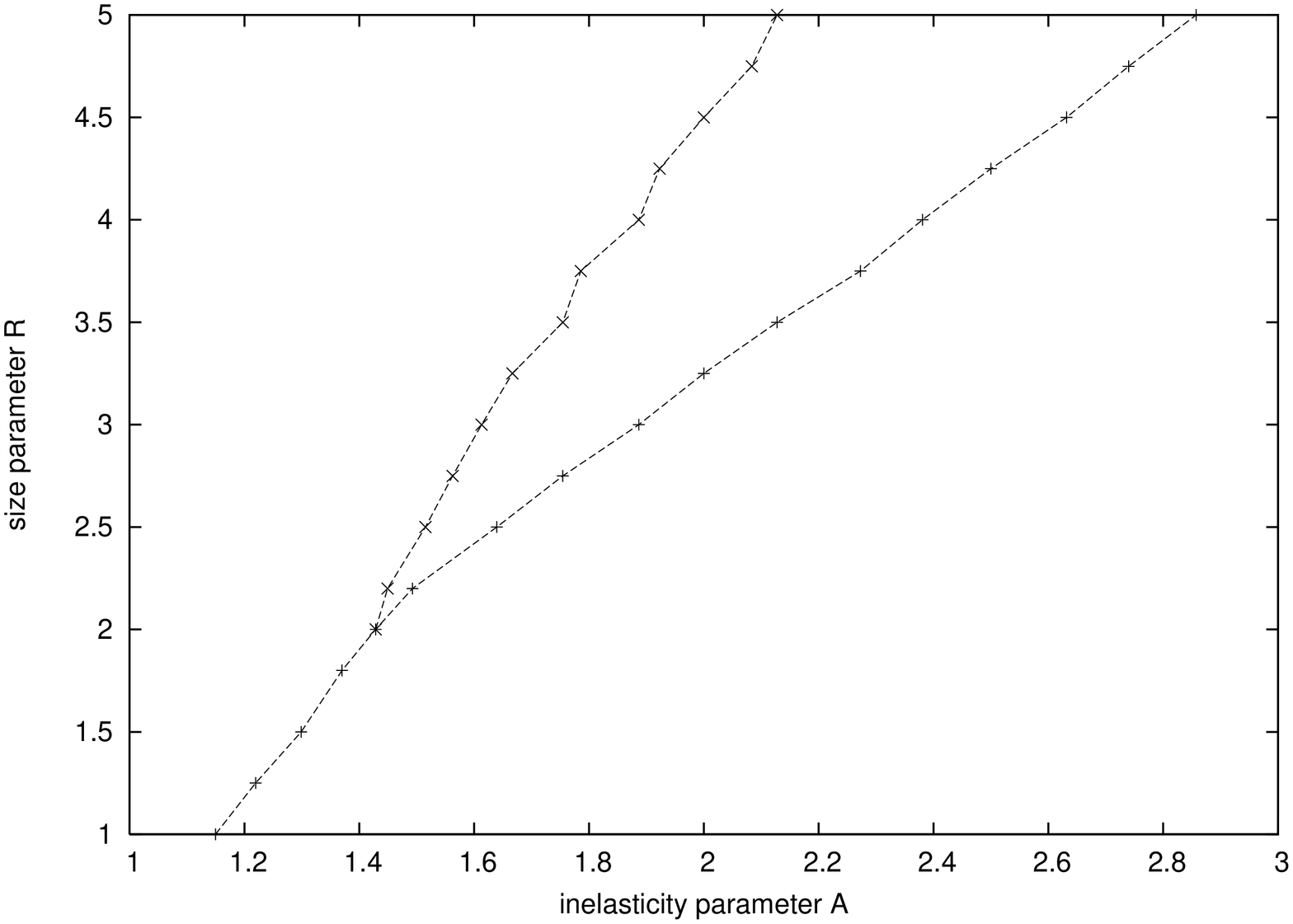}
\caption{\label{figure3} (a) Bifurcation diagram for  the modeling equations (eq. [4]) , 
for $\lambda = 0.1$, R=3.0. (b) Phase diagram, $\lambda = 0.1$.}
\end{figure}

The parameter $\lambda$ in the two last equation introduces the relaxation time  of the segregation mechanism. By construction, the symmetric regime (n=$\frac{1}{2}$, m=$\frac{1}{2}$) is always a solution of the dynamics of the system. The dynamical equations \ref{model} have been solved numerically and they exhibit the three regimes observed by our even driven simulation. In figure \ref{figure3} is presented the bifurcation diagram obtained for the model which as can be seen is very similar to the one obtained by even driven simulations (fig. \ref{figure2b}). It is important to point out that other functional dependences for the escaping rate of the discs have been tested and indeed they did not show qualitative differences. For complementarity, in our analysis we introduced an additive noise into equations  \ref{model} and some features of the simulations that were not observed in the deterministic model are accessed. Namely, the intermittent regime and the presence of satellite peaks around the frequency of the deterministic oscillations.  One must keep in mind that, although the proposed model is relatively simple, it has been formulated by adapting the escaping rate proposed by J. Eggers for a binary mixture of disc. 

\section{Conclusions}

A Maxwell Demon experiment composed by a binary mixture of grains with two different diameters has been studied by means of even-driven simulations. It has been found that the phenomenology of this system is richer than the case of single diameter grain case. In the former case we found a symmetric steady state, an oscillatory regime, an intermittent behavior and a stationary asymmetric regime featuring grain segregation. We have performed simulations over a wide range of parameters values for constructing the bifurcation diagram and for deducing accurately the power spectrum of the system. The observed oscillation phenomenon is mainly due to the combination of two typical granular effects: the vertical segregation with large particles above the smaller ones and the horizontal instability related to the clustering effect.
Moreover, we have proposed a qualitative model for studying binary mixtures of grain inspired 
on the model proposed in \cite{eggers} for the experiment with only one species of discs.
In order to recover the phenomenology obtained by the simulations, we have introduced into our proposition  a segregation mechanism  through two new variables. The numerical integration and the linear stability analysis of our dynamical model (Eq. \ref{model}) showed its excellent ability to reproduce the major dynamical features obtained by simulations, namely; a) transition from a symmetric regime to an oscillation regime (Hopf bifurcation) characterized by a delay between the signal of the small and the large discs, b) passage from the oscillation regime to an asymmetric steady regime. 

It worthwhile to recall that binary demon experiments have been already performed by Van der Weele et al. \cite{deele}. Even when these authors do not observed an oscillatory segregation regime (granular clock) they suggested, based on their experimental results, that the key factor point leading to an horizontal segregation instability may be the difference on the disk radii. In our simulations the radius ratio used was relatively large (8.0 and 4.0). This last may lead to think that the observed dynamics may be restricted to a particular set of parameters. Nonetheless, both our simulations and our modeling equations reproduce undoubtedly the major dynamical features observed by experiments and the results reported by  A. Barrat and  E. Trizac for a binary mixture of grains (without gravity). Recent numerical simulations developed by Marconi \cite{marco} seem to corroborate the existence  of the granular clock existence.  Finally, even when the proposed empirical equations (Eq.\ref{model}) are relatively simple they undoubtedly reproduce the major intriguing features observed experimentally and by simulations. Presently, we are working on a re-formulation of our system of equations (Eq.\ref{model}) which will allow to represent both the horizontal and vertical segregation in such a way that it permits to measure experimentally the parameters associated to the segregation mechanisms. The results of the last approach will be  presented in a forthcoming paper.

\vskip 0.5cm
\centerline{\bf Acknowledgments}
\vskip 0.5cm
We would like to thank Prof. Gregoire Nicolis for the fruitful discussions we had together, M. Mareschal for coining the granular clock. Also we would like to thank the CRI from the Universit\'e de Bourgogne for access to their computer facilities.  R.L. would like to thank U.M.B. Marconi for the enlightening scientific discussions.

\end{document}